\documentclass[prl,twocolumn,aps,letterpaper, preprintnumbers,superscriptaddress]{revtex4-2}
\usepackage{amsthm,amsmath,amssymb}

\usepackage{mathrsfs}
\usepackage{amsthm}
\usepackage{amsmath}
\usepackage{amssymb}
\usepackage{graphicx}
\usepackage{epstopdf}
\usepackage{url}
\usepackage{subfigure}
\usepackage{float}
\usepackage{bm}
\usepackage{ifthen}
\usepackage[usenames,dvipsnames]{color}
\usepackage{mathrsfs}
\usepackage[colorlinks=true,citecolor=blue,urlcolor=black]{hyperref}
\usepackage{float}

\newcommand{\be}{\begin{equation}}
	\newcommand{\ee}{\end{equation}}

\newcommand{\sket}[1]{{\ensuremath{\lvert#1\rangle}}}
\newcommand{\lket}[1]{{\ensuremath{\left\lvert#1\right\rangle}}}
\newcommand{\ket}[1]{\if@display\lket{#1}\else\sket{#1}\fi}

\newcommand{\sbra}[1]{{\ensuremath{\langle#1\rvert}}}
\newcommand{\lbra}[1]{{\ensuremath{\left\langle#1\right\rvert}}}
\newcommand{\bra}[1]{\if@display\lbra{#1}\else\sbra{#1}\fi}

\newcommand{\sbraket}[2]{{\ensuremath{\langle#1\rvert#2\rangle}}}
\newcommand{\lbraket}[2]{{\ensuremath{\left\langle#1\!\left\rvert\vphantom{#1}#2\right.\!\right\rangle}}}
\newcommand{\braket}[2]{\if@display\lbraket{#1}{#2}\else\sbraket{#1}{#2}\fi}

\newcommand{\sketbra}[2]{{\ensuremath{\lvert #1\rangle\!\langle #2\rvert}}}
\newcommand{\lketbra}[2]{{\ensuremath{\left\lvert #1\right\rangle\!\!\left\langle #2\right\rvert}}}
\newcommand{\ketbra}[2]{\if@display\lketbra{#1}{#2}\else\sketbra{#1}{#2}\fi}

\theoremstyle{plain}

\theoremstyle{definition}

\begin{document}

\title{Generalized one-way function and its application}

\author{Hua-Lei Yin}\email{hlyin@ruc.edu.cn}
\affiliation{Department of Physics and Beijing Key Laboratory of Opto-electronic Functional Materials and Micro-nano Devices, Key Laboratory of Quantum State Construction and Manipulation (Ministry of Education), Renmin University of China, Beijing 100872, China}
\date{\today}

\begin{abstract}
One-way functions are fundamental to classical cryptography and their existence remains a longstanding problem in computational complexity theory~\cite{goldreich2006foundations,menezes2018handbook}. Recently, a provable quantum one-way function has been identified, which maintains its one-wayness even with unlimited computational resources~\cite{Yin2024Unconditionally}. Here, we extend the mathematical definition of functions to construct a generalized one-way function by virtually measuring the qubit of provable quantum one-way function and randomly assigning the corresponding measurement outcomes with identical probability. Remarkably, using this generalized one-way function, we have developed an unconditionally secure key distribution protocol based solely on classical data processing, which can then utilized for secure encryption and signature. Our work highlights the importance of information in characterizing quantum systems and the physical significance of the density matrix. We demonstrate that probability theory and randomness are effective tools for countering adversaries with unlimited computational capabilities.
\end{abstract}

\maketitle

%%%%%%%%%%%%%%%%%%%%%%%%%%  body  %%%%%%%%%%%%%%%%%%%%%%%%%%

Cryptography plays a crucial role in protecting personal privacy, safeguarding national security, and advancing the digital economy. Modern cryptography originated from Shannon's introduction of information theory into cryptanalysis, where the one-time pad was demonstrated to offer unconditional security through probability theory~\cite{shannon1949communication}. Specifically, the posterior probability of the plaintext, given the ciphertext in the one-time pad, remains equal to the prior probability of the plaintext, even when adversaries have unlimited computational resources~\cite{shannon1949communication,stinson1995cryptography}. To address key length and key reuse issues, Shannon introduced the concepts of diffusion and confusion~\cite{shannon1949communication}, which facilitated the development of symmetric encryption algorithms such as the Advanced Encryption Standard~\cite{daemen2002design}. To tackle key distribution challenges in large user networks, public-key cryptography was developed~\cite{diffie1976new}. Central to this is the concept of the one-way function, which provides asymmetry~\cite{goldreich2006foundations} and is a critical resource for designing digital signatures~\cite{menezes2018handbook}, zero-knowledge proof~\cite{goldwasser1985knowledge}, and secure multiparty computation~\cite{yao1982protocols}.

One-way function is a particular type of function $f$ that is computationally easy to evaluate in the forward direction but difficult to invert. Specifically, for each input $x$, $f(x)$ can be computed in polynomial time. However, given a random output $f(x)$, it is infeasible to determine the input $x$ using a deterministic Turing machine in polynomial time~\cite{goldreich2006foundations}. Although the existence of one-way functions remains unproven, Numerous public-key cryptographic systems~\cite{diffie1976new,rivest1978method,elgamal1985public,koblitz1987elliptic,miller1985use,bos2018crystals,ducas2018crystals,bernstein2019sphincs} have been proposed based on the assumption of specific one-way functions, with some of these systems evolving into widely adopted standards on the Internet. Indeed, proving the existence of one-way functions would also substantiate the conjecture that $\mathcal{P} \neq \mathcal{NP}$, one of the seven Millennium Prize Problems. However, the truth of $\mathcal{P} \neq \mathcal{NP}$ does not necessarily imply the existence of one-way functions.

Unfortunately, certain problems traditionally considered as one-way functions have been compromised by known quantum algorithms~\cite{shor1999polynomial}. For instance, the quantum Fourier transform can efficiently solve period-finding problems, encompassing prime factorization and discrete logarithm among others. Additionally, the hidden subgroup problem in finite abelian groups is vulnerable to exponential speedup attacks by quantum computations~\cite{mosca1998hidden}.
Consequently, the latest development in public-key cryptography, post-quantum cryptography~\cite{bernstein2017post}, is regarded as resistant to quantum computing attacks~\cite{alagic2022status}. For instance, lattice-based cryptography~\cite{ajtai1996generating,micciancio2009lattice}, which relies on non-commutative hidden subgroup problems, is a prominent example.
Table~\ref{Tab1} summarizes various public-key cryptographic systems and the one-way functions upon which they are based. Actually, even if one-way functions exist, public-key cryptography cannot be secure against adversaries with unlimited computational resources, as such adversaries could potentially exhaustively explore all possible results to find the correct solution. Here, we develop a generalized one-way function with rigorous one-wayness and applied it to the design of unconditionally secure key distribution.

\begin{table*}[t]
\caption{Summary of notable public-key cryptographic systems, outlining the task, one-way function and the
level of security.} \label{Tab1}
\begin{tabular}{c|c|c|c|c}
  \hline \hline
   Scheme & Cryptographic task &  One-way function & Computational security & Unconditional security\\
   & & & (Quantum) &\\
  \hline
  Diffie-Hellman~\cite{diffie1976new} & Key distribution &  Discrete logarithm & No &  No\\
  \hline
  Rivest-Shamir-Adleman~\cite{rivest1978method} & Encryption and signature & Prime factorization & No & No\\
  \hline
  ElGamal~\cite{elgamal1985public} & Encryption and signature & Discrete logarithm & No  & No\\
  \hline
  Elliptic-curve~\cite{koblitz1987elliptic,miller1985use} & Key distribution & Discrete logarithm & No  & No \\
  & Encryption and signature & & &\\
  \hline
  CRYSTALS-Kyber~\cite{bos2018crystals}  & Encryption  & Learning with errors & Maybe & No \\
   \hline
  CRYSTALS-Dilithium~\cite{ducas2018crystals}  & Signature & Learning with errors & Maybe & No \\
    &  & Short integer solution & &\\
  \hline
  SPHINCS$^{+}$~\cite{bernstein2019sphincs} & Signature & Stateless hash function& Maybe & No \\
\hline  \hline
\end{tabular}
\end{table*}

\begin{figure*}[t]
\centering
\includegraphics[width=18cm]{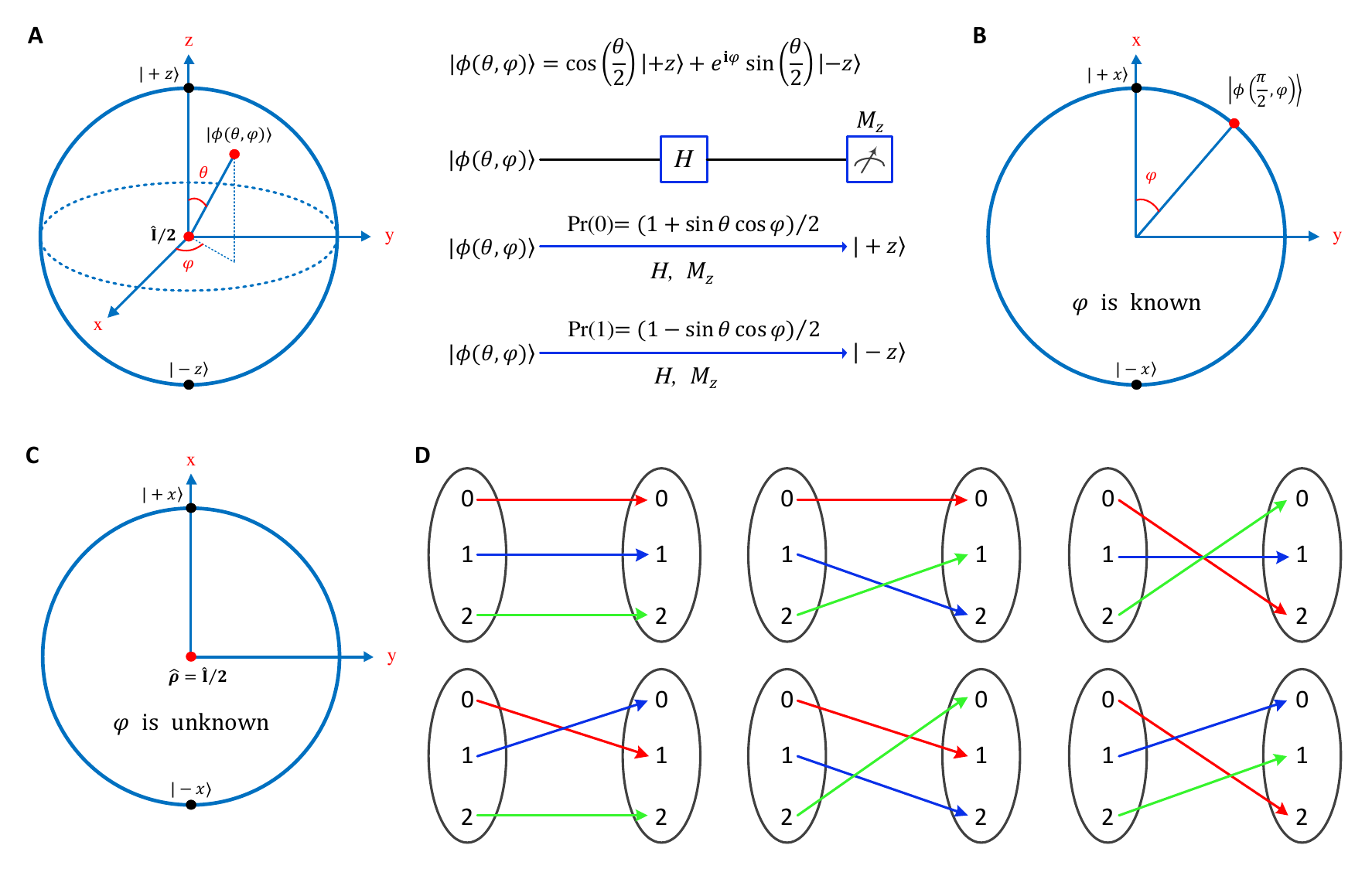}
\caption{\textbf{Pure state and mixed state of a two-dimensional quantum system.} (\textbf{A}) Bloch sphere representation of a density matirx. A superposition state, when measured in the $\mathcal{X}$ basis--which is realized by a Hadamard gate $H$ followed by a $\mathcal{Z}$ basis measurement $M_{z}$--will collapse randomly into one of the eigenstates of the measurement operator with a certain probability. (\textbf{B}) A quantum system can be represented as a pure state $\ket{\phi(\frac{\pi}{2},\varphi)}$ on the periphery of the $\textrm{x}-\textrm{y}$ circle if one has the phase information $\varphi$ and $\theta=\frac{\pi}{2}$. (\textbf{C}) A quantum system can only be represented as the maximally mixed state $\hat{\textbf{I}}/2$ if $\theta=\frac{\theta}{2}$ and no phase information $\varphi$ is available. (\textbf{D}) The random mapping rule $f_{k}:j\rightarrow j'$. There are $m!$ (with $m=3$ as an example) possible random mappings from a finite domain of size $m$ to a finite codomain of size $m$.
} \label{f1}
\end{figure*}

\begin{figure*}[t]
\centering
\includegraphics[width=18cm]{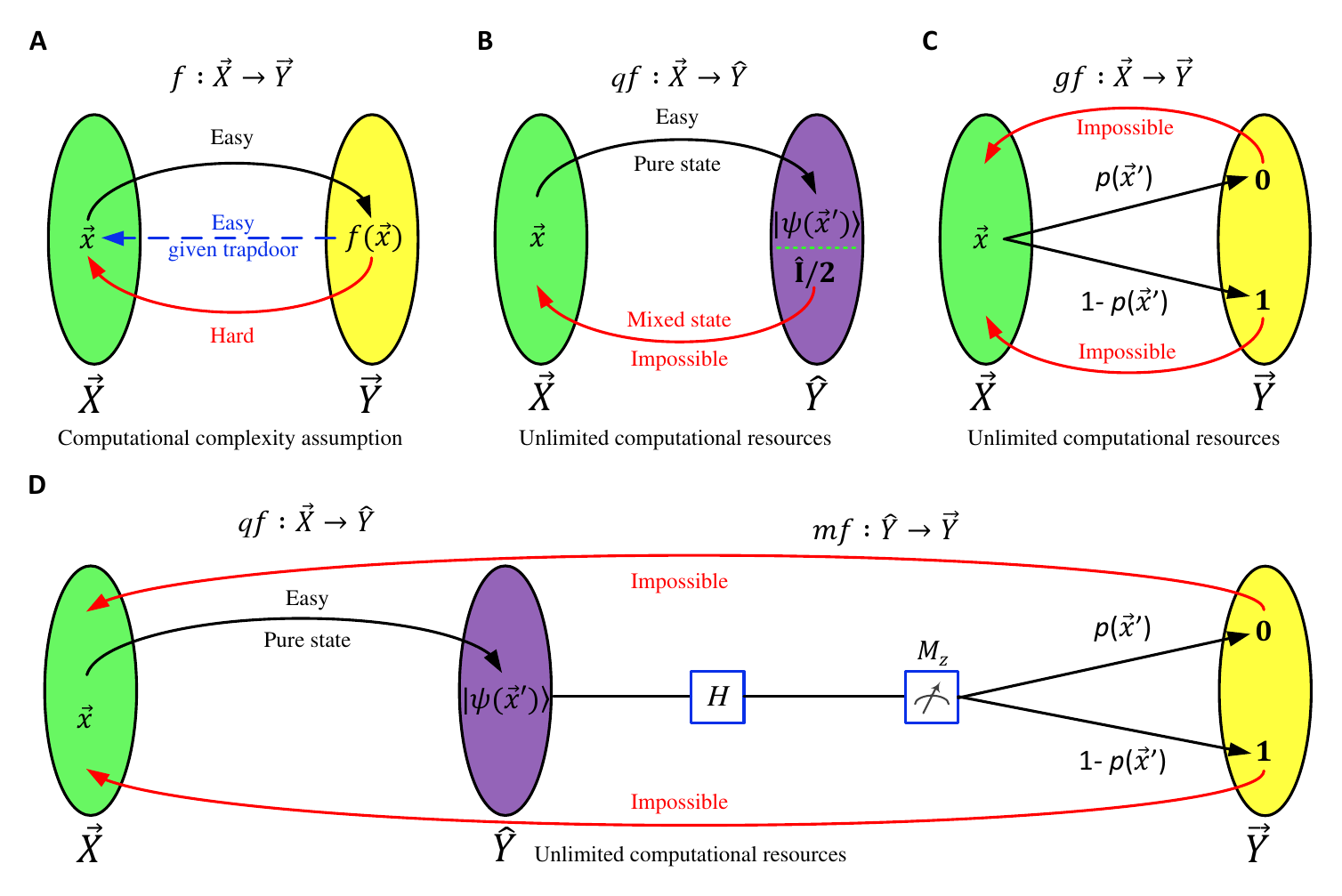}
\caption{\textbf{Comparison of several one-way functions.}  (\textbf{A}) Traditional one-way function based on the computational complexity assumptions.  The forward computation is straightforward, while the reverse problem is very difficult. Once the trapdoor information is acquired, the solution becomes easily accessible. (\textbf{B})  Provable quantum one-way function. The one-wayness is rigorously maintained even with unlimited computational resources. The forward and backward quantum states are entirely distinct, as the quantum state of a system evolves with the information acquired.  (\textbf{C}) Generalized one-way function. (\textbf{D}) Provable quantum one-way function combined with the $\mathcal{X}$ basis measurement. For each input $\vec{x}$ (where the random mapping rule transforms $\vec{x}$ to $\vec{x}'$), there is a probability $p(\vec{x}')$ of obtaining output 0 and $1 - p(\vec{x}')$ of obtaining output 1.
} \label{f2}
\end{figure*}

\subsection{Generalized one-way function}

To introduce our generalized one-way function, let us first retrospect a quantum system fundamentally distinct from its classical counterpart (Fig.~\ref{f1}A). A two-dimensional quantum system can be represented on the Bloch sphere, where a pure state is described by $\ket{\phi(\theta,\varphi)}=\cos\frac{\theta}{2}\ket{+z}+e^{\textbf{i}\varphi}\sin\frac{\theta}{2}\ket{-z}$, lying on the sphere's surface. Here, $\ket{\phi(\theta,\varphi)}$ represents a superposition of $\ket{+z}$ and $\ket{-z}$ with a fixed phase $\varphi$. Measuring in the $\mathcal{X}$ basis yields $\ket{+z}$ with probability $(1+\sin\theta\cos\varphi)/2$ and $\ket{-z}$ with probability $(1-\sin\theta\cos\varphi)/2$. The maximally mixed state $\hat{\textbf{I}}/2$ is located at the center of the Bloch sphere. A quantum system is considered a pure state only if the parameters $\theta$ and $\varphi$ are known. Specifically, $\ket{\phi(\theta=\frac{\pi}{2},\varphi)}=\left(\ket{+z}+e^{\textbf{i}\varphi}\ket{-z}\right)/\sqrt{2}$ lies on the edge of the $\textrm{x}-\textrm{y}$ circle (Fig.~\ref{f1}B). Conversely, if the phase is unknown and random, the system is in a maximally mixed state $\hat{\textbf{I}}/2$ (Fig.~\ref{f1}C). Thus, the quantum state of the same system can vary depending on the observer's information.

In a game between a sender and a receiver, the sender randomly selects one of $m$ symmetric qubits, $\ket{\psi(j)} = \left(\ket{+z} + e^{\textbf{i}\frac{2\pi}{m}j} \ket{-z}\right)/\sqrt{2}$, with equal probability based on a random index $j \in \{0, 1, \ldots, m-1\}$, and sends it to the receiver. The receiver's task is to identify which of the $m$ states was chosen (i.e., determine $j$). From the receiver's perspective, since the phase $\varphi = \frac{2\pi}{m}j$ is completely unknown, the received qubit can only be perceived as a maximally mixed state,
\begin{equation}
\begin{aligned}\label{eq1}
\hat{\rho}=\frac{1}{m}\sum_{j=0}^{m-1}\ket{\psi(j)}\bra{\psi(j)}=\frac{\hat{\textbf{I}}}{2}.
\end{aligned}
\end{equation}
The minimum probability of an incorrect answer by the receiver is $1-2/m$ when the process is repeated independently a sufficient number of times (see supplementary materials). It is close to the error rate of $1-1/m$ that would be achieved by guessing one of the $m$ states completely at random when $m$ is large. To prevent the adversary from inferring fixed bit values based on the phase interval, the concept of a random mapping rule is introduced~\cite{Yin2024Unconditionally}.
\emph{Random Mapping}~\cite{menezes2018handbook}: Let $\mathcal{K}_{m}$ denote the collection of all one-to-one mappings from the domain $\{0, 1, \ldots, m-1\}$ to the codomain $\{0, 1, \ldots, m-1\}$. The $k$-th element of $\mathcal{K}_{m}$ is referred to as a $k$-th random mapping rule $f_{k}: j \rightarrow j'$. Clearly, there are $|\mathcal{K}_{m}| = m!$ possible mapping rules (see Fig.~\ref{f1}D). Assuming that every mapping in $\mathcal{K}_{m}$ is equally likely, it can be determined by random numbers consisting of $m \log_{2} m$ bits~\cite{Yin2024Unconditionally}.
Under the random mapping rule $f_{k}: j \rightarrow j'$, even if the phases of two qubits are nearly identical, their corresponding indices are entirely unrelated.

In public-key cryptography utilizing traditional one-way functions $f:\vec{X}\rightarrow \vec{Y}$ (see Fig.~\ref{f2}A), the output $f(\vec{x})$ derived from the input $\vec{x}$ is classical data. This output is deterministic and accessible to all parties, including potential adversaries. Consequently, the adversary can employ the output $f(\vec{x})$, leveraging unlimited computational resources, to successfully decipher the original input $\vec{x}$.
In contrast, with our provable quantum one-way function $qf:\vec{X}\rightarrow \hat{Y}$ (see Fig.~\ref{f2}B), the output is a quantum system. Due to the application of the random mapping rule $f_{k}:\vec{x}\rightarrow \vec{x}'$ and random input $x$, the adversary lacks any knowledge of the original input $\vec{x}$, even when provided with unlimited computational resources.
From the adversary's perspective, the quantum system cannot be described as a pure state $\ket{\psi(\vec{x}')}$ but must instead be represented by a maximally mixed state $\hat{\textbf{I}}/2$.
This directly provides an asymmetry between the adversary and the legitimate user.
This implies that, regardless of the classical or quantum operations performed by the adversary on the quantum system within the framework of the provable quantum one-way function, the one-wayness remains rigorously maintained.
If we focus solely on the input $\vec{X}$ and output $\vec{Y}$, it becomes evident that the generalized one-way function $gf:\vec{X}\rightarrow \vec{Y}$ (see Fig.\ref{f2}C) is equivalent to the provable quantum one-way function $qf:\vec{X}\rightarrow \hat{Y}$ combined with the $\mathcal{X}$ basis measurement $mf:\hat{Y}\rightarrow \vec{Y}$ (see Fig.\ref{f2}D).
The rigorous one-wayness of the generalized one-way function is preserved because the $\mathcal{X}$ basis measurement can be conceptually regarded as a special quantum operation performed by the adversary in the context of the provable quantum one-way function.

\emph{Generalized One-Way Function}: Consider a set of independent and randomly generated binary bit substrings $\vec{x}_{i} \in \{0,1\}^{\log_{2} m}$, where $i\in\{1,2,\dots,n\}$, which together constitute the input data string $\vec{X} = \vec{x}_{1} || \vec{x}_{2} || \dots || \vec{x}_{n}$. The $k$-th ($k \in \{1,2,\dots,m!\}$) random mapping rule transforms $\vec{x}_{i}$ into $\vec{x}'_{i}$, such that $f_{k}(\vec{x}_{i}) = \vec{x}'_{i}$. The binary bit value $\vec{y}_{i} \in {0,1}$ will form the output data string $\vec{Y} = \vec{y}_{1} || \vec{y}_{2} || \dots || \vec{y}_{n}$. The generalized one-way function $gf:\vec{X} \rightarrow \vec{Y}$ is defined as a multivalued function that maps the input data string $\vec{X} \in \{0,1\}^{n \log_{2} m}$ to the output data string $\vec{Y} \in \{0,1\}^{n}$. The mapping is given by:
\begin{equation}\label{eqs2}
\vec{y}_{i}=gf(\vec{x}_{i})=
   \begin{cases}
   0,~~~\textrm{probability}~\frac{1+\cos\left[\frac{2\pi}{m}f_{k}(\vec{x}_{i})\right]}{2},\\
   \\
   1,~~~\textrm{probability}~\frac{1-\cos\left[\frac{2\pi}{m}f_{k}(\vec{x}_{i})\right]}{2},
   \end{cases}
  \end{equation}
where $m$ is a sufficiently large integer.
Note that the binary bit string is automatically converted to a decimal value as needed during calculations. When preparing $n$ qubit states $\bigotimes_{i=1}^{n}\ket{\psi(\vec{x}'_{i})}$, the output data string $\vec{Y}$ can be directly obtained using the $\mathcal{X}$ basis measurement acting on each qubit state. To introduce an equivalent \emph{virtual measurement} employed in our generalized one-way function, consider the following steps: First, $nb$-bit quantum random numbers are used to form bit string $\vec{A}=\vec{a}_{1}||\vec{a}_{2}||\cdots||\vec{a}_{n}$ with $\vec{a}_{i}\in\{0,1\}^{b}$. Second, if $0\leq \vec{a}_{i}<2^{b-1}\left[1+\cos\left(\frac{2\pi}{m}\vec{x}'_{i}\right)\right]$, then $\vec{y}_{i}=0$; otherwise, $\vec{y}_{i}=1$.

In fact, the rigorous one-wayness of the generalized one-way function can be proven not only using provable quantum one-way function based on density matrix theory, but also through probability theory method.
By calculation (see supplementary materials), we find that
\begin{equation}
\begin{aligned}\label{eq3}
{\rm Pr}[\vec{x}_{i}|\vec{y}_{i}]&=\frac{{\rm Pr}[\vec{y}_{i}|\vec{x}_{i}]{\rm Pr}[\vec{x}_{i}]}{{\rm Pr}[\vec{y}_{i}]}={\rm Pr}[\vec{x}_{i}],\\
\end{aligned}
\end{equation}
and if \(m\) is sufficiently large, the following holds:
\begin{equation}
\begin{aligned}\label{eq4}
{\rm Pr}[\vec{X}|\vec{Y}]&\sim{\rm Pr}[\vec{X}].\\
\end{aligned}
\end{equation}
Thus, the a posteriori probability that the input data is $\vec{X}$, given that the output $\vec{Y}$ is observed, is nearly identical to the prior probability that the input is $\vec{X}$. In other words, after applying the generalized one-way function to a completely random input data string $\vec{X}$, we cannot exclude any possible values of $\vec{X}$; all values of $\vec{X}$ are almost equally probable if we only know the output data string $\vec{Y}$.

\begin{figure*}[t]
\centering
\includegraphics[width=18cm]{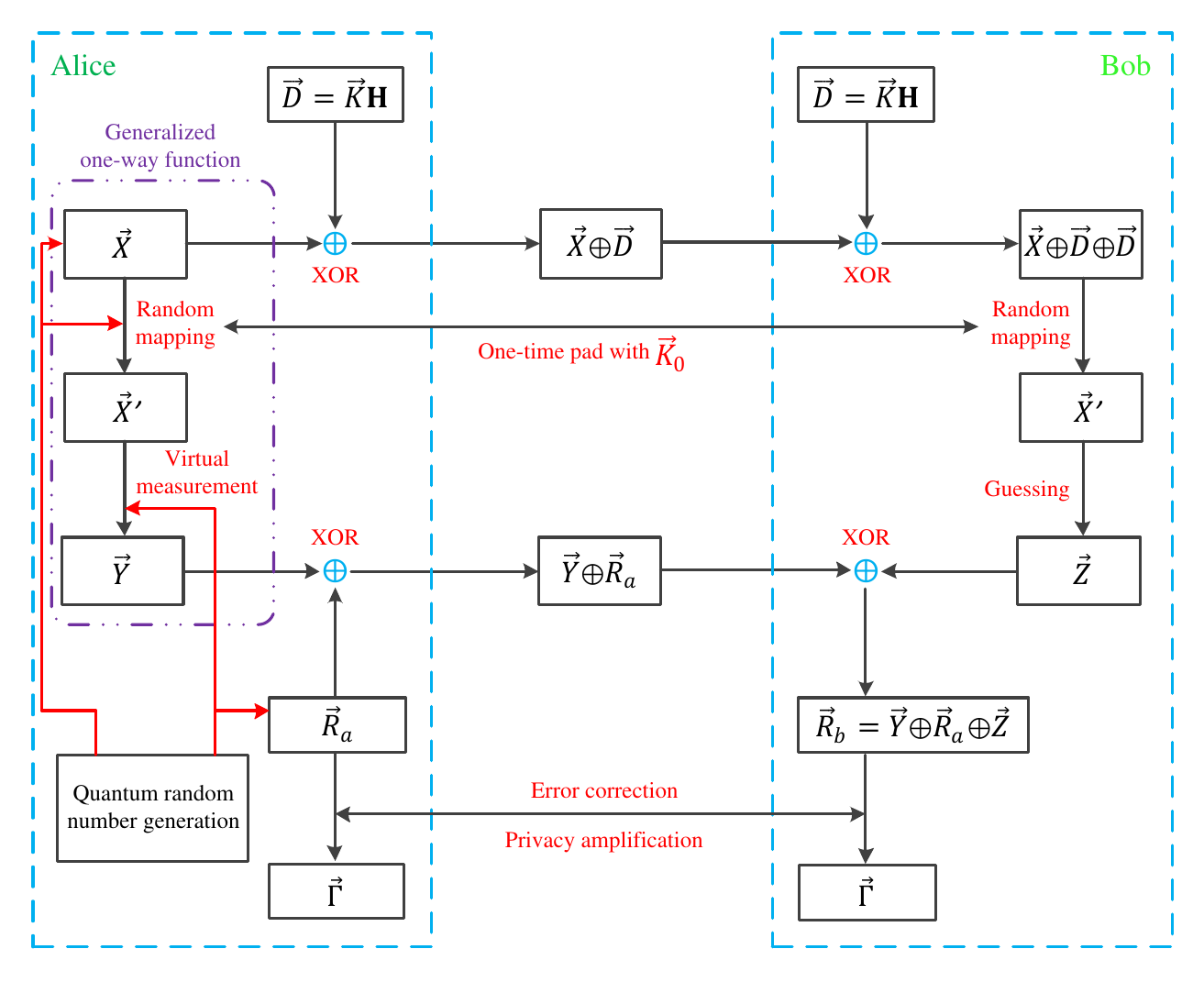}
\caption{\textbf{Schematic diagram of PKD protocol.} In each session, Alice uses the generalized one-way function to generate a bit string $\vec{Y}$ based on the input random bit string $\vec{X}$. Alice and Bob exchange the random bit string $\vec{X}$ using a data string $\vec{D}$, which is produced from the pre-shared bit string $\vec{K}$ and a Toeplitz matrix $\textbf{H}$. Alice then announces the XOR results $\vec{Y} \oplus \vec{R}_{a}$, where $\vec{R}_{a}$ is her raw key. Bob guesses the bit string $\vec{Y}$ to obtain $\vec{Z}$ based on the received $\vec{X}$. To deduce Alice's raw key $\vec{R}_{a}$, Bob computes his own raw key $\vec{R}_{b}$ by performing an XOR operation between $\vec{Y} \oplus \vec{R}_{a}$ and $\vec{Z}$. Finally, Alice and Bob apply error correction and privacy amplification to derive an identical and secret key bit string $\vec{\Gamma}$.
} \label{f3}
\end{figure*}

\subsection{Probability key distribution}

A direct application of the generalized one-way function is the construction of an unconditionally secure key distribution protocol, referred to as Probability Key Distribution (PKD). As illustrated in Fig.~\ref{f3}, our PKD protocol employs full classical data processing, allowing it to be readily implemented in any network environment. A notable feature of PKD is that Alice and Bob are required to share random bit strings $\vec{K} \in \{0,1\}^{s}$, $\vec{K}_{\rm fix} \in \{0,1\}^{s+t-1}$, and $K_{0} \in \{0,1\}^{m\log_{2}^{m}}$. Specifically, the random bit strings $\vec{K}$ and $K_{0}$ are used only once per session and are updated in subsequent sessions, whereas the random bit string $\vec{K}_{\rm fix}$ is reused across thousands of sessions before being updated.

For each session, Alice employs quantum random numbers to decide the $k$-th random mapping rule $f_{k}$~\cite{Yin2024Unconditionally}. She uses the random bit string $\vec{K}_{0}$ as a key and transmits the mapping rule $f_{k}$ to Bob using the one-time pad. Alice utilizes quantum random numbers to construct the input data string $\vec{X}=\vec{x}_{1}||\vec{x}_{2}||\cdots||\vec{x}_{n}\in\{0,1\}^{n\log_{2}^{m}}$, which is then mapped to $\vec{X}'=\vec{x}'_{1}||\vec{x}'_{2}||\cdots||\vec{x}'_{n}$ using the random mapping rule $f_{k}:\vec{x}_{i}\rightarrow\vec{x}'_{i}$. Alice then applies the virtual measurement operation to obtain the output data string $\vec{Y}$.
To meet the criteria of a generalized one-way function $\vec{X}\rightarrow\vec{Y}$, the generated data string $\vec{D}\in\{0,1\}^{n\log_{2}^{m}}$ must cover all possible values in each session, given that Eve has access to $\vec{X}\oplus\vec{D}$. Thus, let $\vec{D}=\vec{K}\textbf{H}$, where $\textbf{H}$ is a Toeplitz matrix with $s$ rows and $t$ columns ($t=n\log_{2}m$), generated from the random bit string $\vec{K}_{\rm fix}$. Due to the rigorous one-wayness of the generalized one-way function, the bit strings $\vec{K}$ and $\vec{K}_{\rm fix}$ remain unknown and random from Eve's perspective, even if Eve knows $\vec{Y}$. Consequently, only Alice and Bob can access the bit string $\vec{X}$. To correctly deduce Alice's raw key $\vec{R}_{a}$, Bob uses the same mapping rule $f_{k}$ to derive $\vec{X}'$ and attempts to estimate $\vec{Y}$ as accurately as possible, resulting in bit string $\vec{Z}=\vec{z}_{1}||\vec{z}_{2}\cdots||\vec{z}_{n}$. One possible approach is to set $\vec{z}_{i}=0$ if $0\leq\vec{x}'_{i}<m/4$ or $3m/4\leq\vec{x}'_{i}<m$, and $\vec{z}_{i}=1$ if $m/4\leq\vec{x}'_{i}<3m/4$.

The value of $\vec{z}_{i}$ is deterministically associated with the interval of $\vec{x}'_{i}$, so Bob cannot disclose any information about his own raw key $\vec{R}_{a}=\vec{R}_{a}\oplus\vec{Y}\oplus\vec{Z}$ throughout the error correction process.
Using Alice's raw key $\vec{R}_{a}$ as a reference, Bob corrects his raw key $\vec{R}_{b}$ to align with Alice's through an error correction algorithm, such as a low-density parity check code. The amount of information required for the error correction step is $\lambda=nfh(E)$. The Shannon entropy function is $h(x)=-x\log_{2}x-(1-x)\log_{2}(1-x)$. $E=\frac{1}{2}-\frac{1}{\pi}\simeq18.2\%$ ($f\geq1$) is the bit error rate (the error correction efficiency) between $\vec{R}_{a}$ and $\vec{R}_{b}$. After error correction, error verification and privacy amplification, if the secret key length $\ell$ of one session satisfies~\cite{Yin2024Unconditionally}
\begin{equation}
\begin{aligned}\label{eq2}
\ell \leq n-\lambda-\log_{2}\frac{2}{\varepsilon_{\rm cor}}-2\log_{2}\frac{3}{2\varepsilon_{\rm sec}},
\end{aligned}
\end{equation}
our PKD protocol is $\varepsilon_{\rm cor}$-correct and $\varepsilon_{\rm sec}$-secret. The net remaining secret key length for each session is $\ell-s-m\log_{2}m$ , as $\vec{K}_{\rm fix}$ is updated only after thousands of sessions, making its cost negligible. The secret key rate of our OKD primarily depends on the data processing rate of system and the quantum random number generation rate. For optional parameters, we set $m=2^{10}$, $n=10^{10}$, $s=10^{4}$, $b=12$, $\varepsilon_{\rm cor}=10^{-15}$ and $\varepsilon_{\rm sec}=10^{-10}$. Actually, if both error correction and error verification communications utilize one-time pad encrypted transmission, one can obtain a nearly perfectly secret key without employing complex privacy amplification post-processing.

On one hand, combining PKD with a one-time pad achieves unconditionally secure encryption, ensuring the confidentiality of information processing~\cite{shannon1949communication}. On the other hand, integrating PKD with one-time universal hashing and secret sharing enables unconditionally secure signatures, providing authenticity, integrity, and non-repudiation in information processing~\cite{yin2023experimental}.

\subsection{Outlook}

In summary, we have developed a generalized one-way function with rigorous one-wayness, addressing a long-standing open problem in computational complexity and cryptography. We have also proposed an unconditionally secure key distribution protocol based on this function, relying entirely on classical data processing. The core of our unconditional security is founded on the randomness of quantum physics, the density matrix theory of quantum mechanics, and probability theory of mathematics. Compared to our recent result~\cite{Yin2024Unconditionally}, this work employs phase-randomized qubit superposition states rather than phase-randomized weak coherent states, which directly translates the quantum state preparation and measurement into a fully classical process. Additionally, the bit error rate of raw keys has improved from $25\%$ to $18.2\%$. These advancements make our PKD protocol suitable for widespread and cost-effective deployment in the emerging digital economy, with high efficiency and unconditional security. We also anticipate that the generalized one-way function will find broad applications in other unconditionally secure cryptographic primitives.

\subsection{Acknowledgements}

We gratefully acknowledge the support from the National Natural Science Foundation of China (No. 12274223).

%%%%%%%%%%%%%%%%%%%%%%% References %%%%%%%%%%%%%%%%%%%%%%%%%
% choose a style
%\bibliographystyle{ieeetr}
%\bibliographystyle{unsrt}
%\bibliographystyle{naturemag}
%\bibliographystyle{apsrev}
%%%%%%%%%%%%%%%%%%%%%%%%%%%%%%%%%%%%%%%
% choose a .bib file
%\bibliography{Bibli}
%%%%%%%%%%%%%%%%%%%%%%%%%%%%%%%%%%%%%%%

\end{document}

\title{Supplementary Information for \\
``Generalized one-way function and its application"}

\author{Hua-Lei Yin}
\email{hlyin@ruc.edu.cn}
\affiliation{Department of Physics and Beijing Key Laboratory of Opto-electronic Functional Materials and Micro-nano Devices, Key Laboratory of Quantum State Construction and Manipulation (Ministry of Education), Renmin University of China, Beijing 100872, China}

\date{\today}

\maketitle

\section{Minimum error discrimination}

For $m$ possible states $\{\hat{\rho}_{j}\}_{j=0}^{m-1}$ with associated a priori probabilities $\{p_{j}\}_{j=0}^{m-1}$, there is a POVM $\{E_{j}\}_{j=0}^{m-1}$ with $m$ elements that can achieve minimum error discrimination.
For many cases, the minimum error discrimination measurement is the square-root measurement. In addition, there is an important conclusion for square-root measurements; i.e., for any set of pure states, there is at least one set of prior probabilities such that the minimum error discrimination measurement for this set of states is the square root measurement.
The POVM elements of the square-root measurement can be given by~\cite{barnett2009quantum}
\begin{equation}
\begin{aligned}
\hat{E}_{j}=p_{j}\hat{\rho}^{-1/2}\hat{\rho}_{j}\hat{\rho}^{-1/2},
\end{aligned}
\end{equation}
where $\hat{\rho}=\sum_{j=0}^{m-1}p_{j}\hat{\rho}_{j}$. Obviously, the above operator $E_{j}$ is positive, and $\sum_{j=0}^{m-1}\hat{E}_{j}=\hat{\textbf{I}}$.

The square-root measurement is the minimum error discrimination measurement for symmetric pure states. Considering $m$ symmetric pure states $\left\{\ket{\psi(j)}=\frac{1}{\sqrt{2}}(\ket{+z}+e^{\textbf{i}2\pi j/m}\ket{-z})\right\}_{j=0}^{m-1}$ with a uniform priori probability of $p_{j}=1/m$. The Gram matrix of the states we are trying to distinguish between is an $m\times m$ matrix, where the matrix element $G_{i,j}$ of the $i$-th row and $j$-th column can be defined as
\begin{equation}
\begin{aligned}
G_{i,j}&=\langle\psi(i)|\psi(j)\rangle=\frac{1}{\sqrt{2}}\left(\bra{+z}+e^{-\textbf{i}2\pi i/m}\bra{-z}\right)\frac{1}{\sqrt{2}}\left(\ket{+z}+e^{\textbf{i}2\pi j/m}\ket{-z}\right)\\
&=\frac{1}{2}\left[1+e^{\textbf{i}2\pi(j-i)/m}\right],
\end{aligned}
\end{equation}
where $i,j=0,1,\ldots,m-1$. Note that we let the matrix start with zero rows and zero columns instead of one row and one column for consistency. Obviously, the Gram matrix $G$ is a circulant matrix since it relies only on the difference $j-i$. It can be diagonalized with the unitary discrete Fourier transform. The eigenvalue $\lambda_{r}$ of Gram matrix $G$ can be given by
\begin{equation}
\begin{aligned}
\lambda_{r}=\sum_{k=0}^{m-1}c_{k}\omega^{kr}
\end{aligned}
\end{equation}
where we have $r=0,1,\ldots,m-1$, $\omega=e^{\textbf{i}2\pi/m}$ and $c_{k}=\frac{1}{2}\left(1+e^{\textbf{i}2\pi k/m}\right)$. The optimal minimum error discrimination probability $P_{\rm min}$ can be written as~\cite{wallden2014minimum}
\begin{equation}
\begin{aligned}
P_{\rm min}&=1-\frac{1}{m^{2}}\left|\sum_{r=0}^{m-1}\sqrt{\lambda_{r}}\right|^{2}\\
&=1-\frac{1}{m^{2}}\left|\sum_{r=0}^{m-1}\sqrt{\sum_{k=0}^{m-1}\frac{e^{\textbf{i}2\pi kr/m}}{2}(1+e^{\textbf{i}2\pi k/m})}\right|^{2},\\
&=1-\frac{2}{m}.
\end{aligned}
\end{equation}

We also have an intuitive explanation of the above minimum error discrimination results $P_{\rm min}=1-2/m$ if $m$ is even. Let two qubit states $\{\ket{\psi(k)},\ket{\psi(k+m/2)}\}$ be the $k$-th set ($0\leq k\leq m/2-1$), there are $m/2$ different sets.
The density matrix of each set is identical to the maximally mixed state, i.e.,
\begin{equation}
\begin{aligned}\label{eq1}
\hat{\rho}_{k}&=\frac{1}{2}\left[\ket{\psi(k)}\bra{\psi(k)}+\ket{\psi(k+m/2)}\bra{\psi(k+m/2)}\right]\\
&=\frac{1}{2}(\ket{+z}\bra{+z}+\ket{-z}\bra{-z})=\frac{\hat{\textbf{I}}}{2}.
\end{aligned}
\end{equation}
Obviously, no one can distinguish the $m/2$ different sets with the same density matrix. The adversary can only randomly guess one set from $m/2$ sets, and the probability of guessing error is $1-2/m$.
Besides, in each set, the two qubit states are orthogonal and thus can be distinguished with 100\% probability. Therefore, the minimum error discrimination probability is $P_{\rm min}=1-\frac{2}{m}$ for $m$ symmetric qubit states.

We remark that $m$ symmetric pure qubit states $\left\{\ket{\psi(j)}\right\}_{j=0}^{m-1}$ are linearly dependent quantum states
\begin{equation}
\begin{aligned}\label{eq1}
\hat{\rho}&=\frac{1}{m}\sum_{j=0}^{m-1}\ket{\psi(j)}\bra{\psi(j)}\\
&=\frac{1}{2}(\ket{+z}\bra{+z}+\ket{-z}\bra{-z})=\frac{\hat{\textbf{I}}}{2},
\end{aligned}
\end{equation}
where unambiguous quantum state discrimination cannot be successfully performed. An unambiguous state discrimination measurement among $m$ linearly dependent qubit states is only possible when at least $m-1$ copies of the states are available~\cite{chefles2001unambiguous}.

If the adversary does not have the quantum state, who can only randomly guess and has an error probability of $1-\frac{1}{m}$. For sufficiently large $m$, the minimum error discrimination is almost the same as a random guess. If we go from guessing the quantum state to guessing the encoded bit substring corresponding to the quantum state, the random mapping rule $f_{k}:j\rightarrow j'$ further removes the difference between the minimum error discrimination measurement and random guessing since each qubit state corresponds to all possible bit substrings.

Consider a quantum system consisting of $n$ subsystems, where each subsystem is one of $m$ symmetric qubits with equal probability. Clearly, the quantum-state sender perceives the entire system as a pure state comprising $n$ qubits. In contrast, from the recipient's perspective, the system appears as a mixed state, represented by $\hat{\rho}^{\otimes n}$, which is the tensor product of $n$ copies of the state $\hat{\rho}$. The mixed state representation $\hat{\rho}^{\otimes n}$ of quantum system can be written as
\begin{equation}
\begin{aligned}\label{eq1}
\hat{\rho}^{\otimes n}&=\frac{\hat{\textbf{I}}^{\otimes n}}{2^{n}}=\left(\frac{1}{m}\sum_{j=0}^{m-1}\ket{\psi(j)}\bra{\psi(j)}\right)^{\otimes n}\\
&=\frac{1}{m^{n}}\left\{\sum_{r_{0}+r_{1}+\cdots+r_{m-1}=n\atop
r_{j}\geq0,~0\leq j\leq m-1} \left[\begin{pmatrix} n  \\ r_{1},r_{2},\cdots,r_{m-1} \end{pmatrix}\bigotimes_{0\leq j\leq m-1}\left(\ket{\psi(j)}\bra{\psi(j)}\right)^{\otimes r_{j}}\right]\right\},
\end{aligned}
\end{equation}
where $r_{j}\geq 0$ is integer and multinomial coefficient is given by
\begin{equation}
\begin{aligned}\label{eq1}
 \begin{pmatrix} n  \\ r_{1},r_{2},\cdots,r_{m-1} \end{pmatrix}=\frac{n!}{\prod_{j=0}^{m-1}r_{j}!}.
\end{aligned}
\end{equation}
Each subsystem is measured independently in the $\mathcal{X}$ basis for the virtual generalized one-way function.
%For sufficiently large $m$, the general joint attack for the sequences of states $\hat{\rho}^{\otimes n}$ is not stronger than the independent attack for state $\hat{\rho}$ because the minimum error discrimination measurement tends to agree with random guesses.

\section{Conditional probability distribution of random variable}

In this section, we first review the generalized one-way function in the main text.
\emph{Generalized one-way function}: Let independent and random binary bit substrings $\vec{x}_{i}\in\{0,1\}^{\log_{2}m}$ with $i=\{1,2,\dots,n\}$ constitute the input data string $\vec{X}=\vec{x}_{1}||\vec{x}_{2}|\dots||\vec{x}_{n}$. The $k$-th ($k\in\{1,2,\dots,m!\}$) random mapping rule makes the $\vec{x}_{i}$ map to $\vec{x}'_{i}$, i.e., $f_{k}( \vec{x}_{i})=\vec{x}'_{i}$.
The binary bit value $\vec{y}_{i}\in\{0,1\}$ will consist the output data string $\vec{Y}=\vec{y}_{1}||\vec{y}_{2}||\dots||\vec{y}_{n}$. Thus, a multivalue function maps the input data string $\vec{X}\in\{0,1\}^{n\log_{2}m}$ to the output data string $\vec{Y}\in\{0,1\}^{n}$ is the generalized one-way function $gf:\vec{X}\rightarrow \vec{Y}$ if we have
\begin{equation}\label{eqs2}
\vec{y}_{i}=gf(\vec{x}_{i})=
   \begin{cases}
   0,~~~\textrm{probability}\frac{1+\cos\left[\frac{2\pi}{m}f_{k}(\vec{x}_{i})\right]}{2},\\
   \\
   1,~~~\textrm{probability}\frac{1-\cos\left[\frac{2\pi}{m}f_{k}(\vec{x}_{i})\right]}{2},
   \end{cases}
  \end{equation}
where $m$ is large enough.

\subsection{Independent attack}

Let us introduce three random variables $\mathcal{X}$, $\mathcal{Y}$ and $\mathcal{K}$. Let $\vec{x}_{i}\in\{0,1\}^{\log_{2}m}$, $\vec{y}_{i}\in\{0,1\}$ and $k\in\{1,2,\ldots,m!\}$ be the values of random variables $\mathcal{X}$, $\mathcal{Y}$ and $\mathcal{K}$, respectively. According to the definition of the generalized one-way function above, the conditional probability distribution of the random variable $\mathcal{Y}$ can be given by
\begin{equation}
\begin{aligned}
{\rm Pr}[\mathcal{Y}=\vec{y}_{i}|(\mathcal{K}=k,\mathcal{X}=\vec{x}_{i})]=\frac{{\rm Pr}(\mathcal{Y}=\vec{y}_{i},\mathcal{K}=k,\mathcal{X}=\vec{x}_{i})}{{\rm Pr}(\mathcal{K}=k,\mathcal{X}=\vec{x}_{i})}=\frac{1+(-1)^{\vec{y}_{i}}\cos\left[\frac{2\pi}{m}f_{k}(\vec{x}_{i})\right]}{2},
\end{aligned}
\end{equation}
where we have probability
\begin{equation}
\begin{aligned}
{\rm Pr}(\mathcal{K}=k,\mathcal{X}&=\vec{x}_{i})=\frac{1}{m!}\frac{1}{m}.\\
\end{aligned}
\end{equation}
Therefore, the joint probability distribution of three random variables $\mathcal{X}$, $\mathcal{Y}$ and $\mathcal{K}$ can be written as
\begin{equation}
\begin{aligned}
{\rm Pr}(\mathcal{Y}=\vec{y}_{i},\mathcal{K}=k,\mathcal{X}=\vec{x}_{i})=\frac{1+(-1)^{\vec{y}_{i}}\cos\left[\frac{2\pi}{m}f_{k}(\vec{x}_{i})\right]}{2m(m!)}.\\
\end{aligned}
\end{equation}
Obviously, for any integer $m\geq2$, one can obtain the following conclusions:
\begin{equation}
\begin{aligned}\label{eq1}
\frac{1}{m}\sum_{j=0}^{m-1}\frac{1\pm\cos[2\pi j/m]}{2}=\frac{1}{2},
\end{aligned}
\end{equation}
and
\begin{equation}
\begin{aligned}\label{eq1}
{\rm Pr}[\mathcal{Y}=\vec{y}_{i}]&=\sum_{k,\vec{x}_{i}}{\rm Pr}[\mathcal{Y}=\vec{y}_{i},\mathcal{K}=k,\mathcal{X}=\vec{x}_{i}]=\frac{1}{2},\\
{\rm Pr}[\mathcal{Y}=\vec{y}_{i}|\mathcal{X}=\vec{x}_{i}]&=\frac{{\rm Pr}(\mathcal{Y}=\vec{y}_{i},\mathcal{X}=\vec{x}_{i})}{{\rm Pr}(\mathcal{X}=\vec{x}_{i})}=\frac{\sum_{k}{\rm Pr}(\mathcal{Y}=\vec{y}_{i},\mathcal{K}=k,\mathcal{X}=\vec{x}_{i})}{{\rm Pr}(\mathcal{X}=\vec{x}_{i})}=\frac{1}{2}.\\
\end{aligned}
\end{equation}
Using Bayes' theorem, we have
\begin{equation}
\begin{aligned}\label{eq1}
{\rm Pr}[\mathcal{X}=\vec{x}_{i}|\mathcal{Y}=\vec{y}_{i}]&=\frac{{\rm Pr}[\mathcal{Y}=\vec{y}_{i}|\mathcal{X}=\vec{x}_{i}]\times{\rm Pr}[\mathcal{X}=\vec{x}_{i}]}{{\rm Pr}[\mathcal{Y}=\vec{y}_{i}]}\\
&={\rm Pr}[\mathcal{X}=\vec{x}_{i}]=\frac{1}{m}.
\end{aligned}
\end{equation}
One can obtain the simplified form ${\rm Pr}[\vec{x}_{i}|\vec{y}_{i}]={\rm Pr}[\vec{x}_{i}]$.
That is, the a posteriori probability $\vec{x}_{i}$, given that the result $\vec{y}_{i}$ is observed, is identical to the a priori probability $\vec{x}_{i}$.
Therefore, if the adversary considers the individual output bits of the generalized one-way function, then all the inputs are equally likely. The adversary cannot obtain any information of the input data $\vec{X}$ from the output data $\vec{Y}$ according to the generalized one-way function.

\subsection{Joint attack}

However, the adversary can also perform a joint analysis of multiple output bits or even entire output bit strings. Here, we provide a direct analysis showing that multi-bit joint analysis approximates single-bit independent analysis in terms of conditional probability and Shannon entropy. For the input data string $\vec{X}=\vec{x}_{1}||\vec{x}_{2}|\dots||\vec{x}_{n}$ and random bit substrings $\vec{x}_{i}\in\{0,1\}^{\log_{2}m}$ with $i=\{1,2,\dots,n\}$, consider a string $\vec{Y}_{d}$ of output bits at any $d$ positions and the corresponding input string $\vec{X}_{d}$, given the independence of random bit substrings $\vec{x}_{i}$. Although the random mapping rule is unknown, it is fixed, meaning that identical input bit strings $\vec{x}_{i}$ will produce identical states.

\subsubsection{Arbitrary two-bit}
Considering that $d=2$ and $i=j,s$, the conditional probability of output $\vec{y}_{j}||\vec{y}_{s}$ given input $\vec{x}_{j}||\vec{x}_{s}$ can be given by
\begin{equation}
\begin{aligned}\label{eq1}
{\rm Pr}[(\vec{y}_{j}||\vec{y}_{s})|(\vec{x}_{j}||\vec{x}_{s})]=\frac{1}{m!}\sum_{k}\frac{1+(-1)^{\vec{y}_{j}}\cos\left[\frac{2\pi}{m}f_{k}(\vec{x}_{j})\right]}{2}\frac{1+(-1)^{\vec{y}_{s}}\cos\left[\frac{2\pi}{m}f_{k}(\vec{x}_{s})\right]}{2}.
\end{aligned}
\end{equation}
If bit substrings $\vec{x}_{j}\neq\vec{x}_{s}$, we have conditional probabilities
\begin{equation}
\begin{aligned}\label{eq1}
{\rm Pr}[(11)|(\vec{x}_{j}||\vec{x}_{s})]={\rm Pr}[(00)|(\vec{x}_{j}||\vec{x}_{s})]&=\frac{1}{m!}\sum_{k}\frac{1+\cos\left[\frac{2\pi}{m}f_{k}(\vec{x}_{j})\right]}{2}\frac{1+\cos\left[\frac{2\pi}{m}f_{k}(\vec{x}_{s})\right]}{2}\\
&=\frac{1}{m(m-1)}\sum_{i=0}^{m-1}\sum_{u=0,u\neq i}^{m-1}\frac{1+\cos\frac{2\pi i}{m}}{2}\frac{1+\cos\frac{2\pi u}{m}}{2}\\
&=\frac{2m-3}{8(m-1)},
\end{aligned}
\end{equation}
and
\begin{equation}
\begin{aligned}\label{eq1}
{\rm Pr}[(10)|(\vec{x}_{j}||\vec{x}_{s})]={\rm Pr}[(01)|(\vec{x}_{j}||\vec{x}_{s})]&=\frac{1}{m!}\sum_{k}\frac{1+\cos\left[\frac{2\pi}{m}f_{k}(\vec{x}_{j})\right]}{2}\frac{1-\cos\left[\frac{2\pi}{m}f_{k}(\vec{x}_{s})\right]}{2}\\
&=\frac{1}{m(m-1)}\sum_{i=0}^{m-1}\sum_{u=0,u\neq i}^{m-1}\frac{1+\cos\frac{2\pi i}{m}}{2}\frac{1-\cos\frac{2\pi u}{m}}{2}\\
&=\frac{2m-1}{8(m-1)}.
\end{aligned}
\end{equation}
If bit substrings $\vec{x}_{j}=\vec{x}_{s}$, we have conditional probabilities
\begin{equation}
\begin{aligned}\label{eq1}
{\rm Pr}[(11)|(\vec{x}_{j}||\vec{x}_{s})]={\rm Pr}[(00)|(\vec{x}_{j}||\vec{x}_{s})]&=\frac{1}{m!}\sum_{k}\frac{1+\cos\left[\frac{2\pi}{m}f_{k}(\vec{x}_{j})\right]}{2}\frac{1+\cos\left[\frac{2\pi}{m}f_{k}(\vec{x}_{s})\right]}{2}\\
&=\frac{1}{m}\sum_{i=0}^{m-1}\frac{1+\cos\frac{2\pi i}{m}}{2}\frac{1+\cos\frac{2\pi i}{m}}{2}\\
&=\frac{3}{8},
\end{aligned}
\end{equation}
and
\begin{equation}
\begin{aligned}\label{eq1}
{\rm Pr}[(10)|(\vec{x}_{j}||\vec{x}_{s})]={\rm Pr}[(01)|(\vec{x}_{j}||\vec{x}_{s})]&=\frac{1}{m!}\sum_{k}\frac{1+\cos\left[\frac{2\pi}{m}f_{k}(\vec{x}_{j})\right]}{2}\frac{1-\cos\left[\frac{2\pi}{m}f_{k}(\vec{x}_{s})\right]}{2}\\
&=\frac{1}{m}\sum_{i=0}^{m-1}\frac{1+\cos\frac{2\pi i}{m}}{2}\frac{1-\cos\frac{2\pi i}{m}}{2}\\
&=\frac{1}{8}.
\end{aligned}
\end{equation}
Obviously, ${\rm Pr}[00]={\rm Pr}[01]={\rm Pr}[10]={\rm Pr}[11]=\frac{1}{4}$. Therefore, the posterior probabilities can be written as
\begin{equation}\label{eq21}
{\rm Pr}[(\vec{x}_{j}||\vec{x}_{s})|(11)]={\rm Pr}[(\vec{x}_{j}||\vec{x}_{s})|(00)]=
   \begin{cases}
   \frac{2m-3}{2(m-1)m^{2}}, ~~~ \vec{x}_{j}\neq\vec{x}_{s},\\
   \\
  \frac{3}{2m^{2}} ,~~~~~~~~~~\vec{x}_{j}=\vec{x}_{s},
   \end{cases}
  \end{equation}
and
\begin{equation}\label{eq22}
{\rm Pr}[(\vec{x}_{j}||\vec{x}_{s})|(10)]={\rm Pr}[(\vec{x}_{j}||\vec{x}_{s})|(01)]=
   \begin{cases}
   \frac{2m-1}{2(m-1)m^{2}}, ~~~ \vec{x}_{j}\neq\vec{x}_{s},\\
   \\
  \frac{1}{2m^{2}} ,~~~~~~~~~~\vec{x}_{j}=\vec{x}_{s}.
   \end{cases}
  \end{equation}
Actually, we have the prior probabilities ${\rm Pr}[\vec{x}_{j}\neq\vec{x}_{s}]=1-\frac{1}{m}$ and ${\rm Pr}[\vec{x}_{j}=\vec{x}_{s}]=\frac{1}{m}$. For a sufficiently large $m$, according to Eqs.~\eqref{eq21} and \eqref{eq22}, we have the conclusion that
\begin{equation}
\begin{aligned}\label{eq1}
{\rm Pr}[(\vec{x}_{j}||\vec{x}_{s})|(\vec{y}_{j}||\vec{y}_{s})]\simeq\frac{1}{m^{2}}={\rm Pr}[\vec{x}_{j}||\vec{x}_{s}].
\end{aligned}
\end{equation}
This implies that the posterior probability of the input data being $\vec{x}_{j}||\vec{x}_{s}$, given the observed output $\vec{y}_{j}||\vec{y}_{s}$, is nearly identical to the prior probability of the input being $\vec{x}_{j}||\vec{x}_{s}$.

Given the output data $\vec{y}_{j}||\vec{y}_{s}$, the Shannon entropy of the input data $\vec{x}_{j}||\vec{x}_{s}$ can be given by
\begin{equation}
\begin{aligned}\label{eq1}
H[(\vec{x}_{j}||\vec{x}_{s})|(00)]=H[(\vec{x}_{j}||\vec{x}_{s})|(11)]&=-\frac{2m-3}{2m}\log_{2}\frac{2m-3}{2(m-1)m^{2}}-\frac{3}{2m}\log_{2}\frac{3}{2m^{2}}\\
&=2\log_{2}m-\left(1-\frac{3}{2m}\right)\log_{2}\frac{2m-3}{2(m-1)}-\frac{3}{2m}\log_{2}\frac{3}{2}\\
&\simeq 2\log_{2}m=H[\vec{x}_{j}||\vec{x}_{s}],\\
H[(\vec{x}_{j}||\vec{x}_{s})|(01)]=H[(\vec{x}_{j}||\vec{x}_{s})|(10)]&=-\frac{2m-1}{2m}\log_{2}\frac{2m-1}{2(m-1)m^{2}}-\frac{1}{2m}\log_{2}\frac{1}{2m^{2}}\\
&=2\log_{2}m-\left(1-\frac{1}{2m}\right)\log_{2}\frac{2m-1}{2(m-1)}+\frac{1}{2m}\\
&\simeq 2\log_{2}m=H[\vec{x}_{j}||\vec{x}_{s}],\\
\end{aligned}
\end{equation}
where we assume that $m$ is big enough.

\subsubsection{Arbitrary three-bit}
Considering $d=3$ and $i=j,s,t$, the probability of output $\vec{y}_{j}||\vec{y}_{s}||\vec{y}_{t}$ given input $\vec{x}_{j}||\vec{x}_{s}||\vec{x}_{t}$ can be given by
\begin{equation}
\begin{aligned}\label{eq1}
{\rm Pr}[(\vec{y}_{j}||\vec{y}_{s}||\vec{y}_{t})|(\vec{x}_{j}||\vec{x}_{s}||\vec{x}_{t})]=\frac{1}{m!}\sum_{k}\frac{1+(-1)^{\vec{y}_{j}}\cos\left[\frac{2\pi}{m}f_{k}(\vec{x}_{j})\right]}{2}\frac{1+(-1)^{\vec{y}_{s}}\cos\left[\frac{2\pi}{m}f_{k}(\vec{x}_{s})\right]}{2}\frac{1+(-1)^{\vec{y}_{t}}\cos\left[\frac{2\pi}{m}f_{k}(\vec{x}_{t})\right]}{2}.
\end{aligned}
\end{equation}
If $\vec{x}_{j}\neq\vec{x}_{s}\neq\vec{x}_{t}$, we have conditional probabilities ${\rm Pr}[(111)|(\vec{x}_{j}||\vec{x}_{s}||\vec{x}_{t})]={\rm Pr}[(000)|(\vec{x}_{j}||\vec{x}_{s}||\vec{x}_{t})]$,
\begin{equation}
\begin{aligned}\label{eq1}
{\rm Pr}[(000)|(\vec{x}_{j}||\vec{x}_{s}||\vec{x}_{t})]&=\frac{1}{m!}\sum_{k}\frac{1+\cos\left[\frac{2\pi}{m}f_{k}(\vec{x}_{j})\right]}{2}\frac{1+\cos\left[\frac{2\pi}{m}f_{k}(\vec{x}_{s})\right]}{2}\frac{1+\cos\left[\frac{2\pi}{m}f_{k}(\vec{x}_{t})\right]}{2}\\
&=\frac{1}{m(m-1)(m-2)}\sum_{i=0}^{m-1}\sum_{u=0,u\neq i}^{m-1}\sum_{v=0,v\neq i,v\neq u}^{m-1}\frac{1+\cos\frac{2\pi i}{m}}{2}\frac{1+\cos\frac{2\pi u}{m}}{2}\frac{1+\cos\frac{2\pi v}{m}}{2}\\
&=\frac{2m^{2}-9m+10}{16(m-1)(m-2)},
\end{aligned}
\end{equation}
and ${\rm Pr}[(001)|(\vec{x}_{j}||\vec{x}_{s}||\vec{x}_{t})]={\rm Pr}[(010)|(\vec{x}_{j}||\vec{x}_{s}||\vec{x}_{t})]={\rm Pr}[(011)|(\vec{x}_{j}||\vec{x}_{s}||\vec{x}_{t})]={\rm Pr}[(100)|(\vec{x}_{j}||\vec{x}_{s}||\vec{x}_{t})]={\rm Pr}[(101)|(\vec{x}_{j}||\vec{x}_{s}||\vec{x}_{t})]={\rm Pr}[(110)|(\vec{x}_{j}||\vec{x}_{s}||\vec{x}_{t})]$,
\begin{equation}
\begin{aligned}\label{eq1}
{\rm Pr}[(001)|(\vec{x}_{j}||\vec{x}_{s}||\vec{x}_{t})]&=\frac{1}{m!}\sum_{k}\frac{1+\cos\left[\frac{2\pi}{m}f_{k}(\vec{x}_{j})\right]}{2}\frac{1+\cos\left[\frac{2\pi}{m}f_{k}(\vec{x}_{s})\right]}{2}\frac{1-\cos\left[\frac{2\pi}{m}f_{k}(\vec{x}_{t})\right]}{2}\\
&=\frac{1}{m(m-1)(m-2)}\sum_{i=0}^{m-1}\sum_{u=0,u\neq i}^{m-1}\sum_{v=0,v\neq i,v\neq u}^{m-1}\frac{1+\cos\frac{2\pi i}{m}}{2}\frac{1+\cos\frac{2\pi u}{m}}{2}\frac{1-\cos\frac{2\pi v}{m}}{2}\\
&=\frac{2m^{2}-5m+2}{16(m-1)(m-2)},
\end{aligned}
\end{equation}
If $\vec{x}_{j}=\vec{x}_{s}\neq\vec{x}_{t}$, we have conditional probabilities ${\rm Pr}[(111)|(\vec{x}_{j}||\vec{x}_{s}||\vec{x}_{t})]={\rm Pr}[(000)|(\vec{x}_{j}||\vec{x}_{s}||\vec{x}_{t})]$,
\begin{equation}
\begin{aligned}\label{eq1}
{\rm Pr}[(000)|(\vec{x}_{j}||\vec{x}_{s}||\vec{x}_{t})]&=\frac{1}{m!}\sum_{k}\frac{1+\cos\left[\frac{2\pi}{m}f_{k}(\vec{x}_{j})\right]}{2}\frac{1+\cos\left[\frac{2\pi}{m}f_{k}(\vec{x}_{s})\right]}{2}\frac{1+\cos\left[\frac{2\pi}{m}f_{k}(\vec{x}_{t})\right]}{2}\\
&=\frac{1}{m(m-1)}\sum_{i=0}^{m-1}\sum_{u=0,u\neq i}^{m-1}\frac{1+\cos\frac{2\pi i}{m}}{2}\frac{1+\cos\frac{2\pi i}{m}}{2}\frac{1+\cos\frac{2\pi u}{m}}{2}\\
&=\frac{3m-5}{16(m-1)},
\end{aligned}
\end{equation}
and ${\rm Pr}[(110)|(\vec{x}_{j}||\vec{x}_{s}||\vec{x}_{t})]={\rm Pr}[(001)|(\vec{x}_{j}||\vec{x}_{s}||\vec{x}_{t})]$,
\begin{equation}
\begin{aligned}\label{eq1}
{\rm Pr}[(001)|(\vec{x}_{j}||\vec{x}_{s}||\vec{x}_{t})]&=\frac{1}{m!}\sum_{k}\frac{1+\cos\left[\frac{2\pi}{m}f_{k}(\vec{x}_{j})\right]}{2}\frac{1+\cos\left[\frac{2\pi}{m}f_{k}(\vec{x}_{s})\right]}{2}\frac{1-\cos\left[\frac{2\pi}{m}f_{k}(\vec{x}_{t})\right]}{2}\\
&=\frac{1}{m(m-1)}\sum_{i=0}^{m-1}\sum_{u=0,u\neq i}^{m-1}\frac{1+\cos\frac{2\pi i}{m}}{2}\frac{1+\cos\frac{2\pi i}{m}}{2}\frac{1-\cos\frac{2\pi u}{m}}{2}\\
&=\frac{3m-1}{16(m-1)},
\end{aligned}
\end{equation}
and ${\rm Pr}[(010)|(\vec{x}_{j}||\vec{x}_{s}||\vec{x}_{t})]={\rm Pr}[(011)|(\vec{x}_{j}||\vec{x}_{s}||\vec{x}_{t})]={\rm Pr}[(100)|(\vec{x}_{j}||\vec{x}_{s}||\vec{x}_{t})]={\rm Pr}[(101)|(\vec{x}_{j}||\vec{x}_{s}||\vec{x}_{t})]$,
\begin{equation}
\begin{aligned}\label{eq1}
{\rm Pr}[(010)|(\vec{x}_{j}||\vec{x}_{s}||\vec{x}_{t})]&=\frac{1}{m!}\sum_{k}\frac{1+\cos\left[\frac{2\pi}{m}f_{k}(\vec{x}_{j})\right]}{2}\frac{1-\cos\left[\frac{2\pi}{m}f_{k}(\vec{x}_{s})\right]}{2}\frac{1+\cos\left[\frac{2\pi}{m}f_{k}(\vec{x}_{t})\right]}{2}\\
&=\frac{1}{m(m-1)}\sum_{i=0}^{m-1}\sum_{u=0,u\neq i}^{m-1}\frac{1+\cos\frac{2\pi i}{m}}{2}\frac{1-\cos\frac{2\pi i}{m}}{2}\frac{1+\cos\frac{2\pi u}{m}}{2}\\
&=\frac{1}{16}.
\end{aligned}
\end{equation}
If $\vec{x}_{j}=\vec{x}_{t}\neq\vec{x}_{s}$, we have conditional probabilities ${\rm Pr}[(111)|(\vec{x}_{j}||\vec{x}_{s}||\vec{x}_{t})]={\rm Pr}[(000)|(\vec{x}_{j}||\vec{x}_{s}||\vec{x}_{t})]$,
\begin{equation}
\begin{aligned}\label{eq1}
{\rm Pr}[(000)|(\vec{x}_{j}||\vec{x}_{s}||\vec{x}_{t})]&=\frac{1}{m!}\sum_{k}\frac{1+\cos\left[\frac{2\pi}{m}f_{k}(\vec{x}_{j})\right]}{2}\frac{1+\cos\left[\frac{2\pi}{m}f_{k}(\vec{x}_{s})\right]}{2}\frac{1+\cos\left[\frac{2\pi}{m}f_{k}(\vec{x}_{t})\right]}{2}\\
&=\frac{1}{m(m-1)}\sum_{i=0}^{m-1}\sum_{u=0,u\neq i}^{m-1}\frac{1+\cos\frac{2\pi i}{m}}{2}\frac{1+\cos\frac{2\pi u}{m}}{2}\frac{1+\cos\frac{2\pi i}{m}}{2}\\
&=\frac{3m-5}{16(m-1)},
\end{aligned}
\end{equation}
and ${\rm Pr}[(101)|(\vec{x}_{j}||\vec{x}_{s}||\vec{x}_{t})]={\rm Pr}[(010)|(\vec{x}_{j}||\vec{x}_{s}||\vec{x}_{t})]$,
\begin{equation}
\begin{aligned}\label{eq1}
{\rm Pr}[(010)|(\vec{x}_{j}||\vec{x}_{s}||\vec{x}_{t})]&=\frac{1}{m!}\sum_{k}\frac{1+\cos\left[\frac{2\pi}{m}f_{k}(\vec{x}_{j})\right]}{2}\frac{1-\cos\left[\frac{2\pi}{m}f_{k}(\vec{x}_{s})\right]}{2}\frac{1+\cos\left[\frac{2\pi}{m}f_{k}(\vec{x}_{t})\right]}{2}\\
&=\frac{1}{m(m-1)}\sum_{i=0}^{m-1}\sum_{u=0,u\neq i}^{m-1}\frac{1+\cos\frac{2\pi i}{m}}{2}\frac{1-\cos\frac{2\pi u}{m}}{2}\frac{1+\cos\frac{2\pi i}{m}}{2}\\
&=\frac{3m-1}{16(m-1)},
\end{aligned}
\end{equation}
and ${\rm Pr}[(001)|(\vec{x}_{j}||\vec{x}_{s}||\vec{x}_{t})]={\rm Pr}[(011)|(\vec{x}_{j}||\vec{x}_{s}||\vec{x}_{t})]={\rm Pr}[(100)|(\vec{x}_{j}||\vec{x}_{s}||\vec{x}_{t})]={\rm Pr}[(110)|(\vec{x}_{j}||\vec{x}_{s}||\vec{x}_{t})]$,
\begin{equation}
\begin{aligned}\label{eq1}
{\rm Pr}[(001)|(\vec{x}_{j}||\vec{x}_{s}||\vec{x}_{t})]&=\frac{1}{m!}\sum_{k}\frac{1+\cos\left[\frac{2\pi}{m}f_{k}(\vec{x}_{j})\right]}{2}\frac{1+\cos\left[\frac{2\pi}{m}f_{k}(\vec{x}_{s})\right]}{2}\frac{1-\cos\left[\frac{2\pi}{m}f_{k}(\vec{x}_{t})\right]}{2}\\
&=\frac{1}{m(m-1)}\sum_{i=0}^{m-1}\sum_{u=0,u\neq i}^{m-1}\frac{1+\cos\frac{2\pi i}{m}}{2}\frac{1+\cos\frac{2\pi u}{m}}{2}\frac{1-\cos\frac{2\pi i}{m}}{2}\\
&=\frac{1}{16}.
\end{aligned}
\end{equation}
If $\vec{x}_{j}\neq\vec{x}_{s}=\vec{x}_{t}$, we have conditional probabilities ${\rm Pr}[(111)|(\vec{x}_{j}||\vec{x}_{s}||\vec{x}_{t})]={\rm Pr}[(000)|(\vec{x}_{j}||\vec{x}_{s}||\vec{x}_{t})]$,
\begin{equation}
\begin{aligned}\label{eq1}
{\rm Pr}[(000)|(\vec{x}_{j}||\vec{x}_{s}||\vec{x}_{t})]&=\frac{1}{m!}\sum_{k}\frac{1+\cos\left[\frac{2\pi}{m}f_{k}(\vec{x}_{j})\right]}{2}\frac{1+\cos\left[\frac{2\pi}{m}f_{k}(\vec{x}_{s})\right]}{2}\frac{1+\cos\left[\frac{2\pi}{m}f_{k}(\vec{x}_{t})\right]}{2}\\
&=\frac{1}{m(m-1)}\sum_{i=0}^{m-1}\sum_{u=0,u\neq i}^{m-1}\frac{1+\cos\frac{2\pi u}{m}}{2}\frac{1+\cos\frac{2\pi i}{m}}{2}\frac{1+\cos\frac{2\pi i}{m}}{2}\\
&=\frac{3m-5}{16(m-1)},
\end{aligned}
\end{equation}
and ${\rm Pr}[(011)|(\vec{x}_{j}||\vec{x}_{s}||\vec{x}_{t})]={\rm Pr}[(100)|(\vec{x}_{j}||\vec{x}_{s}||\vec{x}_{t})]$,
\begin{equation}
\begin{aligned}\label{eq1}
{\rm Pr}[(100)|(\vec{x}_{j}||\vec{x}_{s}||\vec{x}_{t})]&=\frac{1}{m!}\sum_{k}\frac{1-\cos\left[\frac{2\pi}{m}f_{k}(\vec{x}_{j})\right]}{2}\frac{1+\cos\left[\frac{2\pi}{m}f_{k}(\vec{x}_{s})\right]}{2}\frac{1+\cos\left[\frac{2\pi}{m}f_{k}(\vec{x}_{t})\right]}{2}\\
&=\frac{1}{m(m-1)}\sum_{i=0}^{m-1}\sum_{u=0,u\neq i}^{m-1}\frac{1-\cos\frac{2\pi u}{m}}{2}\frac{1+\cos\frac{2\pi i}{m}}{2}\frac{1+\cos\frac{2\pi i}{m}}{2}\\
&=\frac{3m-1}{16(m-1)},
\end{aligned}
\end{equation}
and ${\rm Pr}[(001)|(\vec{x}_{j}||\vec{x}_{s}||\vec{x}_{t})]={\rm Pr}[(011)|(\vec{x}_{j}||\vec{x}_{s}||\vec{x}_{t})]={\rm Pr}[(100)|(\vec{x}_{j}||\vec{x}_{s}||\vec{x}_{t})]={\rm Pr}[(110)|(\vec{x}_{j}||\vec{x}_{s}||\vec{x}_{t})]$,
\begin{equation}
\begin{aligned}\label{eq1}
{\rm Pr}[(001)|(\vec{x}_{j}||\vec{x}_{s}||\vec{x}_{t})]&=\frac{1}{m!}\sum_{k}\frac{1+\cos\left[\frac{2\pi}{m}f_{k}(\vec{x}_{j})\right]}{2}\frac{1+\cos\left[\frac{2\pi}{m}f_{k}(\vec{x}_{s})\right]}{2}\frac{1-\cos\left[\frac{2\pi}{m}f_{k}(\vec{x}_{t})\right]}{2}\\
&=\frac{1}{m(m-1)}\sum_{i=0}^{m-1}\sum_{u=0,u\neq i}^{m-1}\frac{1+\cos\frac{2\pi u}{m}}{2}\frac{1+\cos\frac{2\pi i}{m}}{2}\frac{1-\cos\frac{2\pi i}{m}}{2}\\
&=\frac{1}{16}.
\end{aligned}
\end{equation}
If $\vec{x}_{j}=\vec{x}_{s}=\vec{x}_{t}$, we have conditional probabilities ${\rm Pr}[(111)|(\vec{x}_{j}||\vec{x}_{s}||\vec{x}_{t})]={\rm Pr}[(000)|(\vec{x}_{j}||\vec{x}_{s}||\vec{x}_{t})]$,
\begin{equation}
\begin{aligned}\label{eq1}
{\rm Pr}[(000)|(\vec{x}_{j}||\vec{x}_{s}||\vec{x}_{t})]&=\frac{1}{m!}\sum_{k}\frac{1+\cos\left[\frac{2\pi}{m}f_{k}(\vec{x}_{j})\right]}{2}\frac{1+\cos\left[\frac{2\pi}{m}f_{k}(\vec{x}_{s})\right]}{2}\frac{1+\cos\left[\frac{2\pi}{m}f_{k}(\vec{x}_{t})\right]}{2}\\
&=\frac{1}{m}\sum_{i=0}^{m-1}\frac{1+\cos\frac{2\pi i}{m}}{2}\frac{1+\cos\frac{2\pi i}{m}}{2}\frac{1+\cos\frac{2\pi i}{m}}{2}\\
&=\frac{5}{16},
\end{aligned}
\end{equation}
and ${\rm Pr}[(001)|(\vec{x}_{j}||\vec{x}_{s}||\vec{x}_{t})]={\rm Pr}[(010)|(\vec{x}_{j}||\vec{x}_{s}||\vec{x}_{t})]={\rm Pr}[(011)|(\vec{x}_{j}||\vec{x}_{s}||\vec{x}_{t})]={\rm Pr}[(100)|(\vec{x}_{j}||\vec{x}_{s}||\vec{x}_{t})]={\rm Pr}[(101)|(\vec{x}_{j}||\vec{x}_{s}||\vec{x}_{t})]={\rm Pr}[(110)|(\vec{x}_{j}||\vec{x}_{s}||\vec{x}_{t})]$,
\begin{equation}
\begin{aligned}\label{eq1}
{\rm Pr}[(001)|(\vec{x}_{j}||\vec{x}_{s}||\vec{x}_{t})]&=\frac{1}{m!}\sum_{k}\frac{1+\cos\left[\frac{2\pi}{m}f_{k}(\vec{x}_{j})\right]}{2}\frac{1+\cos\left[\frac{2\pi}{m}f_{k}(\vec{x}_{s})\right]}{2}\frac{1-\cos\left[\frac{2\pi}{m}f_{k}(\vec{x}_{t})\right]}{2}\\
&=\frac{1}{m}\sum_{i=0}^{m-1}\frac{1+\cos\frac{2\pi i}{m}}{2}\frac{1+\cos\frac{2\pi i}{m}}{2}\frac{1-\cos\frac{2\pi i}{m}}{2}\\
&=\frac{1}{16}.
\end{aligned}
\end{equation}
Obviously, ${\rm Pr}[000]={\rm Pr}[001]={\rm Pr}[010]={\rm Pr}[011]={\rm Pr}[100]={\rm Pr}[101]={\rm Pr}[110]={\rm Pr}[111]=\frac{1}{8}$. Therefore, the conditional probabilities can be written as ${\rm Pr}[(\vec{x}_{j}||\vec{x}_{s}||\vec{x}_{t})|(111)]={\rm Pr}[(\vec{x}_{j}||\vec{x}_{s}||\vec{x}_{t})|(000)]$,
\begin{equation}\label{eq1}
{\rm Pr}[(\vec{x}_{j}||\vec{x}_{s}||\vec{x}_{t})|(000)]=
   \begin{cases}
  \frac{2m^{2}-9m+10}{2(m-1)(m-2)m^{3}}, ~~~ \vec{x}_{j}\neq\vec{x}_{s}\neq\vec{x}_{t},\\
   \\
  \frac{3m-5}{2(m-1)m^{3}},~~~~~~~~~~\vec{x}_{j}=\vec{x}_{s}\neq\vec{x}_{t},\\
  \\
  \frac{3m-5}{2(m-1)m^{3}},~~~~~~~~~~\vec{x}_{j}=\vec{x}_{t}\neq\vec{x}_{s},\\
  \\
  \frac{3m-5}{2(m-1)m^{3}},~~~~~~~~~~\vec{x}_{j}\neq\vec{x}_{s}=\vec{x}_{t},\\
  \\
   \frac{5}{2m^{3}} ,~~~~~~~~~~~~~~~~~\vec{x}_{j}=\vec{x}_{s}=\vec{x}_{t},\\
   \end{cases}
  \end{equation}
and ${\rm Pr}[(\vec{x}_{j}||\vec{x}_{s}||\vec{x}_{t})|(001)]={\rm Pr}[(\vec{x}_{j}||\vec{x}_{s}||\vec{x}_{t})|(110)]$,
\begin{equation}\label{eq1}
{\rm Pr}[(\vec{x}_{j}||\vec{x}_{s}||\vec{x}_{t})|(001)]=
   \begin{cases}
  \frac{2m^{2}-5m+2}{2(m-1)(m-2)m^{3}}, ~~~ \vec{x}_{j}\neq\vec{x}_{s}\neq\vec{x}_{t},\\
   \\
  \frac{3m-1}{2(m-1)m^{3}},~~~~~~~~~~\vec{x}_{j}=\vec{x}_{s}\neq\vec{x}_{t},\\
  \\
  \frac{1}{2m^{3}},~~~~~~~~~~~~~~~~~\vec{x}_{j}=\vec{x}_{t}\neq\vec{x}_{s},\\
  \\
  \frac{1}{2m^{3}},~~~~~~~~~~~~~~~~~\vec{x}_{j}\neq\vec{x}_{s}=\vec{x}_{t},\\
  \\
   \frac{1}{2m^{3}} ,~~~~~~~~~~~~~~~~~\vec{x}_{j}=\vec{x}_{s}=\vec{x}_{t},\\
   \end{cases}
  \end{equation}
and ${\rm Pr}[(\vec{x}_{j}||\vec{x}_{s}||\vec{x}_{t})|(010)]={\rm Pr}[(\vec{x}_{j}||\vec{x}_{s}||\vec{x}_{t})|(101)]$,
\begin{equation}\label{eq1}
{\rm Pr}[(\vec{x}_{j}||\vec{x}_{s}||\vec{x}_{t})|(010)]=
   \begin{cases}
  \frac{2m^{2}-5m+2}{2(m-1)(m-2)m^{3}}, ~~~ \vec{x}_{j}\neq\vec{x}_{s}\neq\vec{x}_{t},\\
   \\
  \frac{1}{2m^{3}},~~~~~~~~~~~~~~~~~\vec{x}_{j}=\vec{x}_{s}\neq\vec{x}_{t},\\
  \\
  \frac{3m-1}{2(m-1)m^{3}},~~~~~~~~~~\vec{x}_{j}=\vec{x}_{t}\neq\vec{x}_{s},\\
  \\
  \frac{1}{2m^{3}},~~~~~~~~~~~~~~~~~\vec{x}_{j}\neq\vec{x}_{s}=\vec{x}_{t},\\
  \\
   \frac{1}{2m^{3}} ,~~~~~~~~~~~~~~~~~\vec{x}_{j}=\vec{x}_{s}=\vec{x}_{t},\\
   \end{cases}
  \end{equation}
and ${\rm Pr}[(\vec{x}_{j}||\vec{x}_{s}||\vec{x}_{t})|(011)]={\rm Pr}[(\vec{x}_{j}||\vec{x}_{s}||\vec{x}_{t})|(100)]$,
\begin{equation}\label{eq1}
{\rm Pr}[(\vec{x}_{j}||\vec{x}_{s}||\vec{x}_{t})|(100)]=
   \begin{cases}
  \frac{2m^{2}-5m+2}{2(m-1)(m-2)m^{3}}, ~~~ \vec{x}_{j}\neq\vec{x}_{s}\neq\vec{x}_{t},\\
   \\
  \frac{1}{2m^{3}},~~~~~~~~~~~~~~~~~\vec{x}_{j}=\vec{x}_{s}\neq\vec{x}_{t},\\
  \\
  \frac{1}{2m^{3}},~~~~~~~~~~~~~~~~~\vec{x}_{j}=\vec{x}_{t}\neq\vec{x}_{s},\\
  \\
  \frac{3m-1}{2(m-1)m^{3}},~~~~~~~~~~\vec{x}_{j}\neq\vec{x}_{s}=\vec{x}_{t},\\
  \\
   \frac{1}{2m^{3}} ,~~~~~~~~~~~~~~~~~\vec{x}_{j}=\vec{x}_{s}=\vec{x}_{t}.\\
   \end{cases}
  \end{equation}
Actually, we have the prior probabilities ${\rm Pr}[\vec{x}_{j}\neq\vec{x}_{s}\neq\vec{x}_{t}]=\frac{(m-1)(m-2)}{m^{2}}$ and ${\rm Pr}[\vec{x}_{j}=\vec{x}_{s}\neq\vec{x}_{t}]={\rm Pr}[\vec{x}_{j}=\vec{x}_{t}\neq\vec{x}_{s}]={\rm Pr}[\vec{x}_{j}\neq\vec{x}_{s}=\vec{x}_{t}]=\frac{m-1}{m^{2}}$ and ${\rm Pr}[\vec{x}_{j}=\vec{x}_{s}=\vec{x}_{t}]=\frac{1}{m^{2}}$. For a sufficiently large $m$, we have the conclusion that
\begin{equation}
\begin{aligned}\label{eq1}
{\rm Pr}[(\vec{x}_{j}||\vec{x}_{s}||\vec{x}_{t})|(y_{j}||y_{s}||y_{t})]\simeq\frac{1}{m^{3}}={\rm Pr}[\vec{x}_{j}||\vec{x}_{s}||\vec{x}_{t}].
\end{aligned}
\end{equation}
Given the output data $y_{j}||y_{s}||y_{t}$, the Shannon entropy of the input data can be given by $H[(\vec{x}_{j}||\vec{x}_{s}||\vec{x}_{t})|(000)]=H[(\vec{x}_{j}||\vec{x}_{s}||\vec{x}_{t})|(111)]$,
\begin{equation}
\begin{aligned}\label{eq1}
H[(\vec{x}_{j}||\vec{x}_{s}||\vec{x}_{t})|(000)]&=-\frac{2m^{2}-9m+10}{2m^{2}}\log_{2}\frac{2m^{2}-9m+10}{2(m-1)(m-2)m^{3}}-\frac{3(3m-5)}{2m^{2}}\log_{2}\frac{3m-5}{2(m-1)m^{3}}-\frac{5}{2m^{2}}\log_{2}\frac{5}{2m^{3}}\\
&\simeq 3\log_{2}m=H[\vec{x}_{j}||\vec{x}_{s}||\vec{x}_{t}],
\end{aligned}
\end{equation}
and $H[(\vec{x}_{j}||\vec{x}_{s}||\vec{x}_{t})|(001)]=H[(\vec{x}_{j}||\vec{x}_{s}||\vec{x}_{t})|(010)]=H[(\vec{x}_{j}||\vec{x}_{s}||\vec{x}_{t})|(011)]=H[(\vec{x}_{j}||\vec{x}_{s}||\vec{x}_{t})|(100)]=H[(\vec{x}_{j}||\vec{x}_{s}||\vec{x}_{t})|(101)]=H[(\vec{x}_{j}||\vec{x}_{s}||\vec{x}_{t})|(110)]$,
\begin{equation}
\begin{aligned}\label{eq1}
H[(\vec{x}_{j}||\vec{x}_{s}||\vec{x}_{t})|(001)]&=-\frac{2m^{2}-5m+2}{2m^{2}}\log_{2}\frac{2m^{2}-5m+2}{2(m-1)(m-2)m^{3}}-\frac{(3m-1)}{2m^{2}}\log_{2}\frac{3m-1}{2(m-1)m^{3}}-\frac{2m-1}{2m^{2}}\log_{2}\frac{1}{2m^{3}}\\
&\simeq 3\log_{2}m=H[\vec{x}_{j}||\vec{x}_{s}||\vec{x}_{t}].
\end{aligned}
\end{equation}

\subsubsection{Arbitrary $d$-bit}

From the above calculation, several characteristics can be discerned. First, if the input bit substrings $\vec{x}_{i}$ are not identical, the conditional (posterior) probability suggests that the output string is close to a random guess outcome. Second, if the input bit substrings $\vec{x}_{i}$ are all identical, the conditional (posterior) probability is maximized when the output data string is either all zeros $\vec{y}_{i}=0$ or all ones $\vec{y}_{i}=1$. Third, if the input bit substrings $\vec{x}_{i}$ are identical, the conditional (posterior) probability is minimized when the output data string contains an equal number of zeros and ones.

For $d$ identical input bit substrings $\vec{x}_{i}=\vec{x}$, the maximum conditional probability ${\rm Pr}[(11\cdots1)|(\vec{x}||\vec{x}||\cdots||\vec{x})]={\rm Pr}[(00\cdots0)|(\vec{x}||\vec{x}||\cdots||\vec{x})]$ can written as
\begin{equation}
\begin{aligned}\label{eq1}
{\rm Pr}[(00\cdots0)|(\vec{x}||\vec{x}||\cdots||\vec{x})]&=\frac{1}{m!}\sum_{k}\left\{\frac{1+\cos\left[\frac{2\pi}{m}f_{k}(\vec{x})\right]}{2}\right\}^{d}\\
&=\frac{1}{m}\sum_{i=0}^{m-1}\left(\frac{1+\cos\frac{2\pi i}{m}}{2}\right)^{d}\\
&=\frac{1}{2\pi}\int_{0}^{2\pi}\left(\frac{1+\cos t}{2}\right)^{d}dt\\
&={_2}{F}_{1}\left(\frac{1}{2},-2d;1;2\right)\leq\frac{1}{2},
\end{aligned}
\end{equation}
where ${_2}{F}_{1}\left(a,b;c;z\right)$ is the Gaussian hypergeometric function and we utilized the conversion between integration and summation.

For $d$ identical input bit substrings $\vec{x}_{i}=\vec{x}$, the minimum conditional probability is that the number of bits of 0 in the output string is the closest to the number of bits of 1.
If $d$ is even, the minimum conditional probability can be given by
\begin{equation}
\begin{aligned}\label{eq1}
{\rm Pr}[(00\cdots0||11\cdots1)|(\vec{x}||\vec{x}||\cdots||\vec{x})]&=\frac{1}{m!}\sum_{k}\left\{\frac{1+\cos\left[\frac{2\pi}{m}f_{k}(\vec{x})\right]}{2}\right\}^{d/2}\left\{\frac{1-\cos\left[\frac{2\pi}{m}f_{k}(\vec{x})\right]}{2}\right\}^{d/2}\\
&=\frac{1}{m}\sum_{i=0}^{m-1}\left(\frac{1+\cos\frac{2\pi i}{m}}{2}\right)^{d/2}\left(\frac{1-\cos\frac{2\pi i}{m}}{2}\right)^{d/2}\\
&=\frac{1}{2\pi}\int_{0}^{2\pi}\left(\frac{1+\cos t}{2}\right)^{d/2}\left(\frac{1-\cos t}{2}\right)^{d/2}dt\\
&=\frac{d!}{2^{2d}\left(\frac{d}{2}!\right)^{2}}\leq\frac{1}{2^{2d-1}}.
\end{aligned}
\end{equation}
If $d$ is odd, the minimum conditional probability can be given by
\begin{equation}
\begin{aligned}\label{eq1}
{\rm Pr}[(00\cdots0||11\cdots1)|(\vec{x}||\vec{x}||\cdots||\vec{x})]&=\frac{1}{m!}\sum_{k}\left\{\frac{1+\cos\left[\frac{2\pi}{m}f_{k}(\vec{x})\right]}{2}\right\}^{(d+1)/2}\left\{\frac{1-\cos\left[\frac{2\pi}{m}f_{k}(\vec{x})\right]}{2}\right\}^{(d-1)/2}\\
&=\frac{1}{m}\sum_{i=0}^{m-1}\left(\frac{1+\cos\frac{2\pi i}{m}}{2}\right)^{(d+1)/2}\left(\frac{1-\cos\frac{2\pi i}{m}}{2}\right)^{(d-1)/2}\\
&=\frac{1}{2\pi}\int_{0}^{2\pi}\left(\frac{1+\cos t}{2}\right)^{(d+1)/2}\left(\frac{1-\cos t}{2}\right)^{(d-1)/2}dt\\
&=\frac{(d-1)!}{2^{2d-1}\left(\frac{d-1}{2}!\right)^{2}}\leq\frac{1}{2^{2d-1}}.
\end{aligned}
\end{equation}

Through straightforward analysis, we find that the conditional probability of the output bit string corresponding to the other input bit string cases must be between the maximum and minimum values of the above identical input bit substrings case.
Therefore, we obtain the following inequality:
\begin{equation}
\begin{aligned}\label{eq1}
\frac{1}{2^{2d-1}}\leq{\rm Pr}[\vec{Y}_{d}|\vec{X}_{d}]\leq\frac{1}{2}.
\end{aligned}
\end{equation}
Note that, the prior probability that all $d$ input bit substrings $\vec{x}_{i}$ are identical is $m^{-d}$, which is very small when both $m$ and $d$ are large. In many instances, the conditional probability closely approximates the probability of a random guess, that is, ${\rm Pr}[\vec{Y}_{d}|\vec{X}_{d}] \approx 2^{-d}$.

According to Bayes' theorem, we have
\begin{equation}
\begin{cases}\label{eq1}
{\rm Pr}[\vec{X}_{d}|\vec{Y}_{d}]&=\frac{{\rm Pr}[\vec{Y}_{d}|\vec{X}_{d}]{\rm Pr}[\vec{X}_{d}]}{{\rm Pr}[\vec{Y}_{d}]}
\leq\frac{\frac{1}{2}m^{-d}}{2^{-d}}=\frac{2^{d-1}}{m^{d}}\simeq\frac{1}{m^{d}}={\rm Pr}[\vec{X}_{d}],\\
\\
{\rm Pr}[\vec{X}_{d}|\vec{Y}_{d}]&=\frac{{\rm Pr}[\vec{Y}_{d}|\vec{X}_{d}]{\rm Pr}[\vec{X}_{d}]}{{\rm Pr}[\vec{Y}_{d}]}
\geq\frac{2^{-2d+1}m^{-d}}{2^{-d}}=\frac{2^{-d+1}}{m^{d}}\simeq\frac{1}{m^{d}}={\rm Pr}[\vec{X}_{d}],\\
\end{cases}
\end{equation}
where we assume that $m$ is sufficiently large, for example, $m=2^{10}$. Therefore, we conclude that ${\rm Pr}[\vec{X}|\vec{Y}]\sim{\rm Pr}[\vec{X}]$.

%%%%%%%%%%%%%%%%%%%%%%%%%%%%%%%%%%%%%%%%
% choose a style
%\bibliographystyle{ieeetr}
%\bibliographystyle{unsrt}
%\bibliographystyle{naturemag}
%\bibliographystyle{apsrev4-1}
%%%%%%%%%%%%%%%%%%%%%%%%%%%%%%%%%%%%%%%

%%%%%%%%%%%%%%%%%%%%%%%%%%%%%%%%%%%%%%%%
% choose a .bib file
%\bibliography{Bibli}
%%%%%%%%%%%%%%%%%%%%%%%%%%%%%%%%%%%%%%%%